\documentstyle[12pt,aasms4]{article}

\newcommand{\V}{$V$~}
\newcommand{\I}{$I$~}
\newcommand{\etal}{$et \; al.$~}

\def\lesssim{\mathrel{\hbox{\rlap{\hbox{%
 \lower4pt\hbox{$\sim$}}}\hbox{$<$}}}}
\def\gtrsim{\mathrel{\hbox{\rlap{\hbox{%
 \lower4pt\hbox{$\sim$}}}\hbox{$>$}}}}

\def\arcmin{\hbox{$^\prime \;$}}

\def\farcs{\hbox{$.\!\!^{\prime\prime}$}}

\begin{document}

\title{A Lack of Planets in 47 Tucanae from an HST Search
\altaffilmark{1}}

\altaffiltext{1}{Based on observations with the NASA/ESA {\it Hubble
Space Telescope} obtained at ST ScI,
which is operated by AURA, Inc. under NASA contract NAS 5-26555.}
\author{Ronald L. Gilliland,
\altaffilmark{2}
T. M. Brown,
\altaffilmark{3}
P. Guhathakurta,
\altaffilmark{4}
A. Sarajedini,
\altaffilmark{5}
E. F. Milone,
\altaffilmark{6}
M. D. Albrow,
\altaffilmark{2}
N. R. Baliber,
\altaffilmark{7}
H. Bruntt,
\altaffilmark{8}
A. Burrows,
\altaffilmark{9}
D. Charbonneau,
\altaffilmark{3,10}
P. Choi,
\altaffilmark{4}
W. D. Cochran,
\altaffilmark{7}
P. D. Edmonds,
\altaffilmark{10}
S. Frandsen,
\altaffilmark{8}
J. H. Howell,
\altaffilmark{4}
D. N. C. Lin,
\altaffilmark{4}
G. W. Marcy,
\altaffilmark{11}
M. Mayor,
\altaffilmark{12}
D. Naef,
\altaffilmark{12}
S. Sigurdsson,
\altaffilmark{13}
C. R. Stagg,
\altaffilmark{6}
D. A. VandenBerg,
\altaffilmark{14}
S. S. Vogt,
\altaffilmark{4}
and M. D. Williams.
\altaffilmark{6}}

\altaffiltext{2}{ST ScI, 3700 San Martin Drive,
Baltimore, MD 21218, gillil@stsci.edu}
\altaffiltext{3}{High Altitude Observatory, National Center for Atmospheric
Research, P.O. Box 3000, Boulder, CO 80307.  NCAR is sponsored by
the National Science Foundation.}
\altaffiltext{4}{UCO/Lick Observatory, University of California at Santa Cruz,
Santa Cruz, CA 95064.}
\altaffiltext{5}{Astronomy Department, Wesleyan University, Middletown, 
CT 06459.}
\altaffiltext{6}{Physics and Astronomy Dept., University of Calgary,
Calgary, T2N 1N4, Canada.}
\altaffiltext{7}{University of Texas at Austin, Department of Astronomy,
Austin, TX 78712.}
\altaffiltext{8}{Institute for Physics and Astronomy, Aarhus University,
DK-8000, Aarhus C, Denmark.}
\altaffiltext{9}{Department of Astronomy, U. of Arizona, 933 N.
Cherry Avenue, Tucson, AZ 85721.}
\altaffiltext{10}{Harvard-Smithsonian Center for Astrophysics, 60 Garden Street,
Cambridge, MA 02138.}
\altaffiltext{11}{Department of Astronomy, University of California, Berkeley, 
CA 94720.}
\altaffiltext{12}{Observatoire de Gen\`{e}ve, CH-1290 Sauverny,
Switzerland.}
\altaffiltext{13}{Dept. Astro., Penn State University,
University Park, PA 16802.}
\altaffiltext{14}{Dept. of Physics and Astronomy, U. of Victoria,
Victoria, BC, V8W 3P6, Canada.}

\begin{abstract}
We report results from a large {\em Hubble Space Telescope} project to
observe a significant ($\sim$34,000) ensemble of main sequence stars
in the globular cluster 47 Tucanae with a goal of defining the frequency
of inner-orbit, gas-giant planets.
Simulations based on the characteristics of the 8.3 days of time-series data
in the F555W and F814W WFPC2 filters show that $\sim$17
planets should be detected by photometric transit signals if the
frequency of hot Jupiters found in the solar neighborhood is assumed to hold for 47 Tuc.
The experiment provided high-quality data sufficient to detect planets.
A full analysis of these WFPC2 data reveals $\sim$75 variables, but
{\bf no} light curves resulted for which a convincing interpretation as
a planet could be made.
The planet frequency in 47 Tuc is at least an order of magnitude below that
for the solar neighborhood.
The cause of the absence of close-in planets in 47 Tuc is not yet known;
presumably the low metallicity and/or crowding of 47 Tuc interfered with planet
formation, with orbital evolution to close-in positions, or with planet 
survival.  
\end{abstract}


\section{Introduction}

The discovery of 51 Peg b (Mayor and Queloz 1995) in a remarkably tight orbit
of 4.2 days around its Sun-like host star challenged prevailing theoretical
views and impelled rapid progress in expanded radial velocity (RV) surveys
which have now resulted in about 50 planet detections;
see recent review by Marcy, Cochran and Mayor (2000).
The existence of inner-orbit, gas-giant planets (hot Jupiters)
enables a highly-efficient photometric search for planets, since
with tight orbits (0.04 -- 0.05 AU) hot Jupiters present
about a 10\% chance of transiting the host star given random orbital inclinations.
Indeed, the transiting planet of HD 209458 was observed (Charbonneau \etal 2000,
Henry \etal 2000) ``on schedule", just as the ensemble of RV-detected hot Jupiters
neared the point of a 50\% expectation for transits.
HD 209458b verified the theory of Guillot \etal (1996) predicting somewhat
extended radii of $\sim$ 1.2 -- 1.4 $R_J$ as a result of retarded cooling due to
high irradiance from the host star.
A transit depth of 1.7\% and duration of 3.0 hrs for HD 209458b repeating
every 3.525 days presents an enticing photometric signal.

In 1998 when the number of detected extra-solar planets was $\sim$10,
of which 4 were hot Jupiters (defined herein as $P_{orb} \, < $ 5 days),
we proposed (in the ``scientifically risky" category) to use WFPC2 on {\em HST}
to observe a large ensemble of stars in 47 Tuc.
Our target was selected: (1) to provide an ideal spatial and brightness 
distribution of stars matching $HST$ capabilities, and (2) to shed insight
into understanding the origins of planets by observing a system with 
reduced metallicity [Fe/H] = $-$0.7, [$\alpha$/Fe] = +0.4 dex (Salaris and Weiss 1998).
Our proposed field, with the crowded core of 47 Tuc on the PC1 CCD of WFPC2, provides
some 40,000 main sequence targets (giants are not of interest since the transit
depth is equal to $(R_P / R_* )^2$,
where $R_P$, $R_*$ are the planet and star radii respectively).
Saturation on the bright end occurs near the cluster turnoff where stellar 
radii are rapidly increasing and thus the expected signal dropping anyway.
Stars far down ($\sim$4 mags) the main sequence remain viable targets
since a rising signal from falling stellar radii would balance declining
signal-to-noise (S/N) for these fainter stars.
The frequency of hot Jupiters (9 are now known) in the solar neighborhood
is about 1\%, with about a 10\% chance per system of random orbital 
inclinations yielding transits; the 47 Tuc ensemble should thus provide on the order of
one planet per 1000 surveyed stars.
\section{Observations}

To detect two consecutive transits requires an observing interval
twice the orbital period with continuous (relative to $\sim$2 -- 3 hr transit 
timescale) coverage.
Our 8.3 days (120 orbits of {\em HST}, GO-8267) of continuous observation spanned
1999 July 3 -- 11, the only data gaps resulted from Earth occultations and 
passages through the South Atlantic Anomaly (SAA).
We chose the F555W and F814W filters for the primary time series with 160 s
exposures (cycled every 4 min in each), yielding saturation near \V $\sim$
17.1 at cluster turnoff.
These two filters were alternated in a sequence typically consisting of 6$\times$F555W
and 6$\times$F814W during each SAA free orbit.
Visibility periods impacted by the SAA were alternately devoted to one or the
other filter.
With the orbital period of $HST \,\sim$96.4 min, each transiting system
of interest should display at least two consecutive transits over the 8.3 days in each of 
two filters.
(Transits should be gray, while chance superposition of a main sequence star
and a large amplitude but faint eclipsing binary, which in superposition mimics a planet transit
in depth, would display very red signals.)
The number of 160 s exposures obtained was: 636 (F555W),
653 (F814W).
We took care to 
design the observations with significant margin such that minor changes
in one or even several assumptions regarding number of stars, realized
photometric precision, planet radius and hence signal amplitude, or assumed
frequency of systems would not jeopardize a robust result.

The primary consideration in detection margin is the ratio of signal 
amplitude to the time series precision multiplied by (for Gaussian noise) the square root of the 
number of data points during transits.
Time series precisions reach nearly the Poisson limit, {\em e.g.,} at 
\V = 18.4 in 160 s a S/N of 200 (or 0.0053 mag $rms$) results in the F555W filter.
The length of transits (including reduction by $\pi$/4 for average chord lengths)
in hours (geometry plus Kepler's 3rd law)
and probability in \% of transits per existing system given random inclinations are:
\begin{equation}
\tau_{tran} = 1.412 M_*^{-1/3} R_* P_{orb}^{1/3} \; \; {\rm{and}}, \; \; Pr_{tran} = 23.8 M_*^{-1/3} R_* P_{orb}^{-2/3},
\end{equation}
\noindent
where $M_*$ and $R_*$ are in solar units and $P_{orb}$ in days.
For a 47 Tuc star at \V = 18.4, $M_*$ = 0.81, $R_*$ = 0.92 (Bergbusch \& VandenBerg 1992),  $\tau_{tran}$ = 2.20
hrs (up to 2.80 for central passages)
for $P_{orb}$ = 3.8 d
and $Pr_{tran}$ = 9.6\%.
The transit depth (assuming $R_P$ = 1.3$R_J$) is predicted to be 0.022 (about 
4$\sigma$ per observation) yielding a 11.5$\sigma$ detection per transit per filter.
Overall this example would provide about a 23$\sigma$ detection.
\section{Analyses and Time Series Examples}

Full discussion of the analysis steps will appear in Gilliland \etal (2000);
here we provide only a sketch of steps relevant to reaching Poisson noise
limited results for under-sampled, dithered, crowded-field photometry
with the added complication of frame-to-frame focus drift.
We do not believe that any software packages previously described in the
literature would be up to the specialized and stringent requirements of 
this project and have therefore developed our own procedures and codes.
Gilliland \etal (1995) discuss issues relevant
to precise time-series photometry for under-sampled and dithered $HST$
data; the steps to robust cosmic-ray elimination using a polynomial
expansion of the detected intensity as a function of $x,y$ offsets to 
create an over-sampled mean model remain central to the processing 
(Gilliland, Nugent \& Phillips 1999 provide further details).

Initial frame-to-frame offsets are determined by PSF fits to 4 relatively
isolated stars in each frame.
For each pixel a ``surface" fit of intensity as a function of sub-pixel
$x,y$ offsets is formed with iterative elimination of multi-sigma
positive deviations (cosmic rays).
The frame-to-frame offsets are then adjusted by solving for the $x,y$
offsets in a least-squares sense that provide a best match for evaluation
of the analytic model (using only pixels on bright, unsaturated stars) to each frame.
The image model and registration (including plate scale changes) solution
are iterated two or three times to convergence.

Absolute photometry has been performed on $\times$4 over-sampled, co-added images 
using DAOPHOT II (Stetson 1992).
A typical PSF FWHM is 1.4 pixels; PC and WF CCD plate scales are 0\farcs046
and 0\farcs1 per direct pixel respectively.
Figure 1 shows the CMD of 47 Tuc developed from our deep, well-dithered, 
co-added images in F555W (101,760s) and F814W (104,480s).
Zero points of the instrumental magnitudes have been adjusted to best match
47 Tuc $V$, \I fiducials (Kaluzny \etal 1998) near main sequence turnoff.
Stars have been excluded if: 1) nominal apertures (69 pixel PC, 45 pixel WF)
touch saturated pixels from neighbors, 2) apertures include any bad pixels
flagged in the data quality files (except saturation), 3) if $>$ 90\% of the
light in the aperture comes from wings of brighter nearby neighbors, or 4)
more than 1\% of the frames in both F555W or F814W show saturation in the 
core (gain = 14 $\rm e^- /DN$).
For the results discussed further below only the 34,091 stars falling within 
a bright main sequence box as shown were analyzed for time series.

The state-of-the-art for crowded-field, time-series photometry now involves
creation of difference images ({\em e.g.,} Alcock \etal 1999;
Alard 1999), where for well-sampled, ground-based CCD data excellent gains
over classical PSF fitting in direct images are realized.
With good difference images non-variable objects are removed (except for
residual, unavoidable Poisson noise) leaving any variables clearly present
as isolated (positive or negative) PSFs even if the variable was badly
blended with brighter stars in the direct images.
Extraction of precise relative photometry changes for any star in a difference
image can be handled with either aperture photometry or PSF fitting,
and precise knowledge of the PSF is much less critical for the difference
images relative to attempting photometry on blended stars in the direct image.

Difference images were created for each individual frame using the intensity
of $x,y$ analytic model evaluated at the position of each individual frame.
Such images, however, showed large residuals at each star in many images as
a result of focus changes.
The 2$\mu$ ``breathing" of $HST$ due to changing thermal stresses over orbits
leads to a full amplitude redistribution of light inside-to-outside of a 1 pixel
radius of $\sim$20\%.
We have added the extra step of solving for a compensation kernal such that when convolved
onto the analytic $I(x,y)$ model it best matches individual frames.   
An iteration is now adopted over this focus compensation, the registration
solution and the $x,y$ intensity analytic model.
Difference images can now be formed accounting for $x,y$ offsets and focus
changes; variable stars become obvious in movies of
such images that were invisible in direct image movies.

A time series was created by fitting a PSF
at the known position
of individual stars in each image.
Normalization was performed using counts as a function of magnitude based on
archival calibration images of the standard GRW+70D5824.
Each time series was cleaned by removing any changes linearly correlated with
an ensemble average term and nine vectors comprising all terms through cubic
in $x,y$ offsets.
This decorrelation step usually does not provide much change for the inherently
excellent $HST$ time series.

Figure 2 shows the resulting time series $rms$ of relative intensities 
(magnitudes would be larger by $\times$1.086) for each of 34,091 stars in
the F555W filter (F814W results are similar).
The lower curve shows the ideal photometry limit:  our results are usually
within $\sim$10 -- 15\% of this.
The upper ``Detection limit" curve is a function of assumed planet radius,
stellar radius along the main sequence, length of transits and density of sampling
such that a (Gaussian noise) $rms$ value falling below this curve would 
provide $>$ 6.5$\sigma$ detection for two transits in each filter.
The steep dropoff of the detection capability on the bright end results from
increasing stellar radii near turnoff creating a lower predicted transit signal.
At the faint end smaller stellar radii produce deeper, but shorter transits
the S/N for which is overtaken by the loss in time-series S/N in the sky
background limited regime.

Figure 3 shows a time series that presents a close approximation
to expectations for a planet transit.  
The star has \V = 19.01, \V -- \I = 0.89 and two neighbors brighter by
$\sim$2.5 magnitudes at 0\farcs5 separation.
Decreases of $\sim$3\% can be seen repeating every 1.34 days in both the
direct $V$- and $I$-band time series (a combined 28 $\sigma$ detection).
Predicted transit depths from a planet at 1.3$R_J$ are 0.030 as seen.
A 4$\sigma$ significant secondary eclipse at phase 0.5 suffices for
arguing this is not a planet.
Moreover, analysis of the saturated neighboring stars shows that one
is a large amplitude eclipsing binary at the same period and phase.
The light curve plotted results from the eclipsing-binary PSF wings at
the position of the faint star providing a diluted signal of an ordinary
variable; this is not a planet candidate.

\section{Transit Search and Detection Efficiency}

To search for multiple transits, we fold the time series of each individual
star in each filter with sufficient trial phases and periods within 0.5 to 8.3 
days to densely cover phase space.
We then convolve the folded time series with the theoretical light curve
corresponding to nominal transits.
The convolution is normalized so that, for a white noise input, the 
convolution values are normally distributed with unit variance.
Possible transits are indicated by period--phase combinations with large
positive values of the convolution (see Fig. 3).
Since the noise in the $HST$ data is close to white, we selected a threshold
of 6.3$\sigma$ which should yield $\lesssim$ 1 false alarm for the
entire search space given Gaussian statistics (e.g. Bevington 1969).
A detailed description of procedures and results will appear in Brown \etal (2000).

A key issue in interpreting detections (or lack thereof) is the efficiency with
which real transits would be detected by our search algorithms.
To estimate this efficiency we inserted artificial transits into the data 
stream and processed the resulting modified data sets using the standard
transit-detection pipeline.
These were blind tests, in which the analyst had no knowledge of the properties
of the inserted transits, nor of which stars were affected.
One set of simulations inserted artificial transits by manipulating pixel values
in the original images associated with 24 stars.
By modifying the data at the image level, we verified that extensive image reduction
and time series extraction steps used did not damage real transit signals.
Our procedures correctly retained evidence of transits through
the image processing and time series levels of the analysis.

A more extensive test inserted artificial transits directly into the time
series of about 10000 randomly chosen stars.
This is enough samples to estimate the dependence of our detection efficiency
on planetary radius and orbital period for the actual noise characteristics of
our time series as a function of $V$.
Figure 4 shows that the percentage of correctly categorized transit light 
curves depends strongly on assumed planet radius and more moderately on 
orbital period.
The theoretical expectation (Guillot \etal 1996) is that (irradiated) planet radii vary
only slowly as a function of mass.  Indeed, planets with sub-jovian mass may
have slightly larger radii than a 1$M_J$ planet.
Also, over the two orders of magnitude mass range from planets to brown dwarfs
to stars at the main sequence threshold, radii are not expected to vary by
more than 10 -- 20\%.
This lack of radius dependence on mass would be a hindrance if trying to 
interpret a handful of weak signals as planets (rather than stars),
but for a null result it implies a lack in the sample of any transiting planets,
brown dwarfs or very late M dwarfs.

Assuming an occurrence rate for close-in giant planets that is the same as
in the solar neighborhood (0.8 -- 1.0\%) and a 10\% probability of favorable 
orbital alignment, there should be about 30 transiting planets among our
sample of 34091 stars.
Assuming planet radii of 1.3$R_J$ and a typical period of 3.5 days, and 
allowing for the actual distribution of \V magnitudes (stellar radii and
time series noise) in our sample, the number of planets actually detected
should have been about 17.
Since we saw none, we may conclude (with very high confidence) that in so far
as giant planet occurrence is concerned, the solar neighborhood and 47 Tuc
represent different populations.

\section{Astrophysical Interpretation}

It was noted (Gonzalez 1997) with the first planet detections that host 
stars tend to be considerably more metal rich than the average star
surveyed in the solar neighborhood.
This correlation has been maintained (Laughlin 2000) with a much larger
sample now available.
A correlation with metallicity could, however, be either cause or effect.
It could be that lower metallicity in protoplanetary nebulae causes
a lower frequency of planet formation as a result of fewer dust grains for
nucleation.  It could be that higher metallicity in systems with a 
remaining close-in planet is an effect of inward migration of metal-rich
planets onto the star (Lin, Bodenheimer \& Richardson 1996) thus polluting
the thin outer convective layer.
Our 47 Tuc results are at least consistent with the hypothesis that lower
metallicity biases against formation of hot Jupiters.

47 Tucanae is a massive cluster with a density $\sim 10^3 M_{\odot}\rm{pc}^{-3}$
at 1\arcmin from the core (a typical location for our observations).
In such a crowded environment, close encounters can result in dynamical
interactions, particularly with passing binary systems, which lead to large
changes in orbital parameters (Heggie 1975).
Sigurdsson (1992) considered the orbital stability problem for
high-mass-ratio systems (in the context of pulsar-planets) specifically for 47 Tuc
conditions and concluded that orbits as short as 5 days would be quite
stable against disruption or forced merger even in the most dense core region.
An avenue for destroying hot Jupiters in the crowded 47 Tuc core arises from
consideration of tidal dissipation (Sigurdsson, Lin \& Gilliland 2000) within
the planet.
In this scenario star-planet encounters with binaries can induce an 
eccentric planet orbit.
Dissipation of the stellar tidal disturbance within the planets drains their
orbital energy.
Internal heating may cause the planets to expand and provide positive feedback
to a disruptive process.
Critical to this scenario is the actual frequency of close binaries in 
47 Tuc which will be a valuable side-product of our survey for planets
(Albrow \etal 2000).
Alternative scenarios have been postulated where crowding limits planet 
formation (Armitage 2000).

We have shown that planets like 51 Peg b that are found in $\sim$1\% of local
stars surveyed must be an order of magnitude rarer in the lower-metallicity,
crowded-stellar environment extant in the center of 47 Tucanae.
This represents a significant result delineating where planets exist.
Further observations of stars in different circumstances will be necessary
to learn whether the dominant influence in reducing the planet frequency is low metallicity, a 
crowded environment, or some combination of these or other factors.

\acknowledgments
We thank Merle Reinhart and Patricia Royle at STScI for assistance in
scheduling these unique observations.
This work was supported in part by STScI Grant GO-8267.01-97A to
the Space Telescope Science Institute and several STScI grants
from the same proposal to co-I institutions.

\clearpage
\begin{figure}
\caption{Color magnitude diagram of 46,422 stars from all four WFPC2 CCDs.
The box along the main sequence over 17.1 $<$ \V $<$ 21.6 shows the selection
domain for the 34,091 stars reported herein (PC1 box extended to \V = 16.1).
Numerical entries provide predicted transit depths in magnitudes and duration
(central passage)
in hours assuming a $P_{orb}$ = 3.8 day planet with $R \, = \, 1.3 R_J$.}

\caption{Final time series quality shown as standard deviation of intensity
changes divided by the mean intensity for each star in the WFPC2
F555W bandpass.  The lower curve defines the expected precision
limit based on Poisson noise for isolated stars and background plus readout noise.
Deviations to higher values usually follow either from real variables, or 
increased (background) Poisson noise from near neighbors.
The upper ``Detection limit" curve defines the line below which transits 
are predicted to provide $>$ 6.5$\sigma$ detection for two transits in each
filter -- at this threshold no false alarms should arise.}

\caption{From top: direct \V \& \I light curves; same phased at 1.340 days
after smoothing over 0.036 bins (error bars show standard error from
scatter within bins). 
Transit depths are as expected for a 1.3$R_J$ planet and 0.77$R_{\odot}$
star.
Demonstrates excellent sensitivity to transits, however as discussed in
in the text this is a diluted eclipsing-binary signal -- not a planet.
The bottom panels show the transit detection statistic described in \S 4.}

\caption{Left panel shows detection frequency averaged over periods from simulations as a function
of assumed planet size; right panel shows same as a function of orbital 
period at $R \,=\, 1.2R_{Jup}$.}

\end{figure}

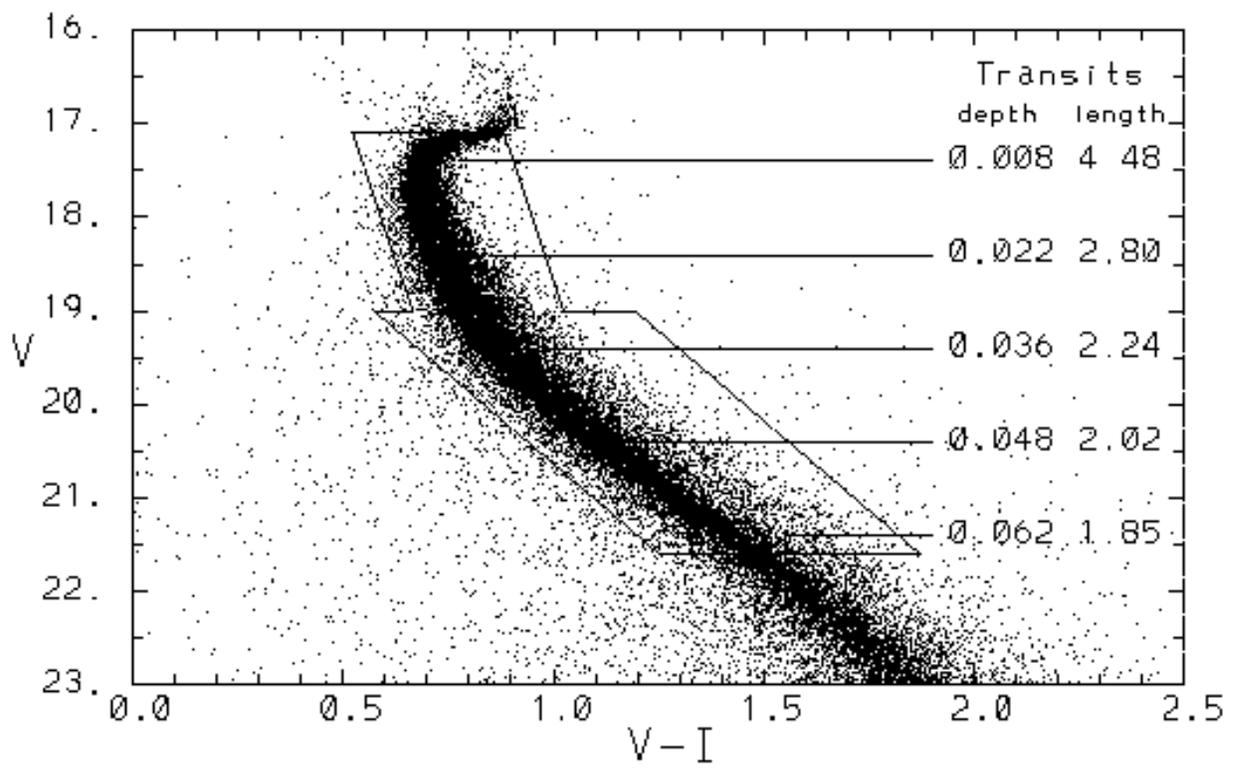
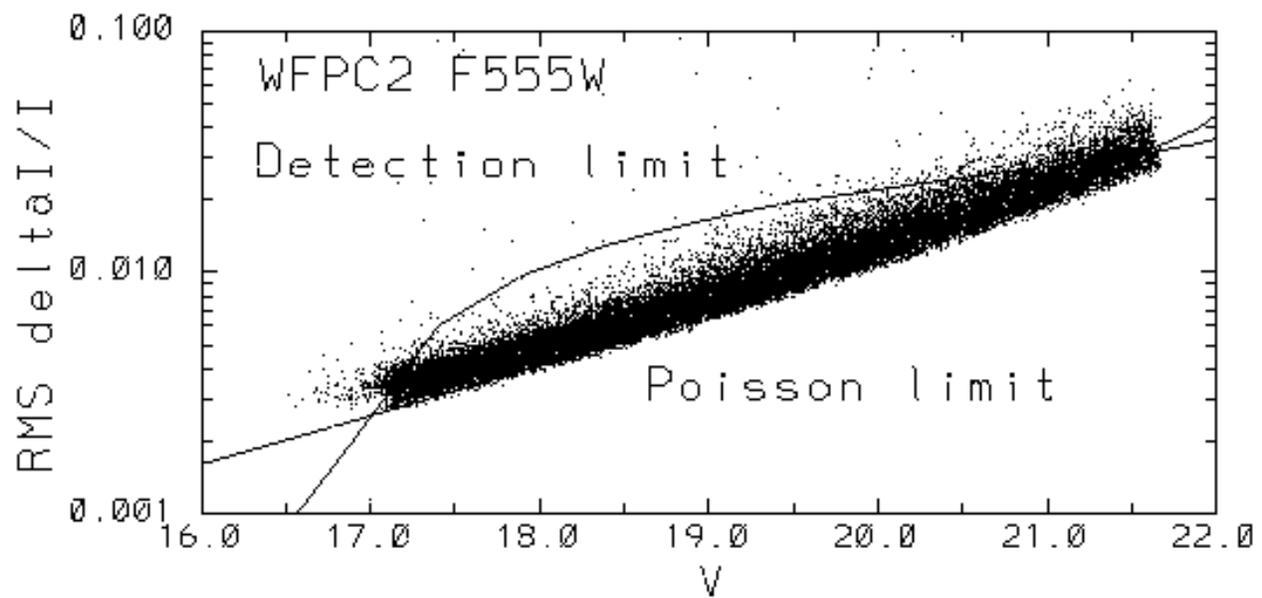
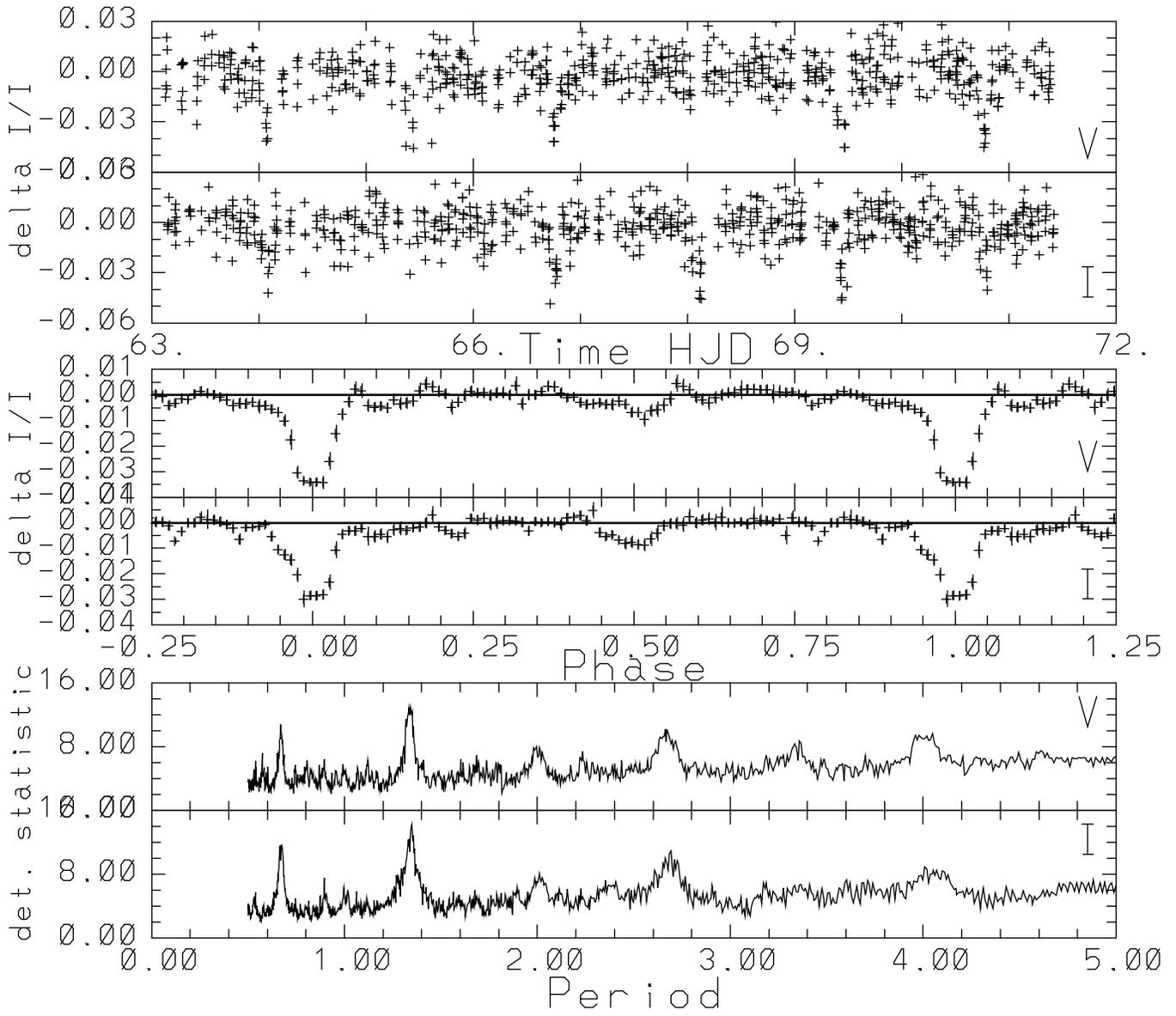
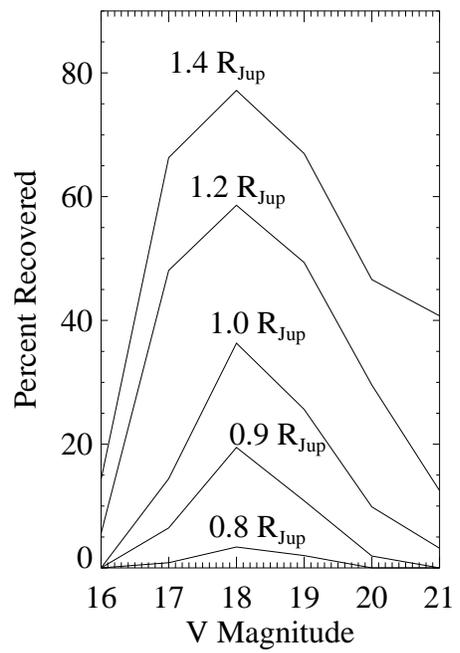
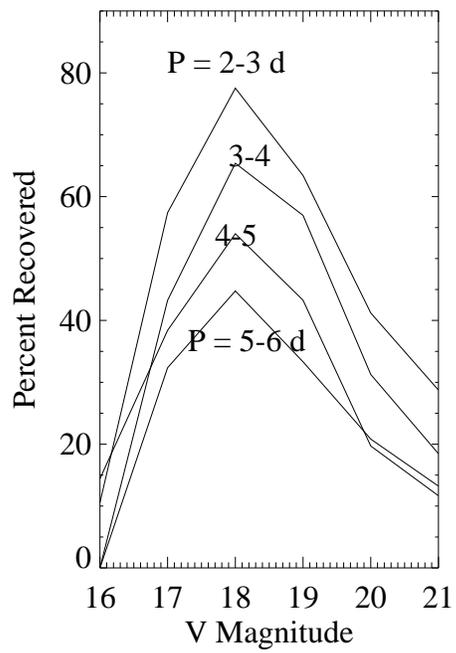
\end{document}